\documentclass[letterpaper]{article} % DO NOT CHANGE THIS
\usepackage[]{aaai2026}  % DO NOT CHANGE THIS 
\usepackage{times}  % DO NOT CHANGE THIS
\usepackage{helvet}  % DO NOT CHANGE THIS
\usepackage{courier}  % DO NOT CHANGE THIS
\usepackage[hyphens]{url}  % DO NOT CHANGE THIS
\usepackage{graphicx} % DO NOT CHANGE THIS
\usepackage{natbib}  % DO NOT CHANGE THIS AND DO NOT ADD ANY OPTIONS TO IT
\usepackage{caption} % DO NOT CHANGE THIS AND DO NOT ADD ANY OPTIONS TO IT
\usepackage{cite}
\usepackage{amsmath,amssymb,amsfonts}

\usepackage{textcomp}
\usepackage{multirow}
\usepackage{multicol}
\usepackage{booktabs}
\usepackage{makecell}
\usepackage{xcolor}
\usepackage{algorithm}
\usepackage{algorithmic}
\usepackage{newfloat}
\usepackage{listings}
\DeclareCaptionStyle{ruled}{labelfont=normalfont,labelsep=colon,strut=off} % DO NOT CHANGE THIS
\floatstyle{ruled}
\newfloat{listing}{tb}{lst}{}
\floatname{listing}{Listing}
\title{EndoIR: Degradation-Agnostic All-in-One Endoscopic Image Restoration via Noise-Aware Routing Diffusion}
\author{
    Tong Chen\textsuperscript{\rm 1}\equalcontrib,
    Xinyu Ma\textsuperscript{\rm 2}\equalcontrib, Long Bai\textsuperscript{\rm 3}\equalcontrib,
    Wenyang Wang\textsuperscript{\rm 1},
    Yue Sun\textsuperscript{\rm 2}\thanks{Co-Corresponding Author},
    Luping Zhou\textsuperscript{\rm 1}\thanks{Corresponding Author}
}
\affiliations{
    \textsuperscript{\rm 1}School of Electrical and Computer Engineering, The University of Sydney, Sydney, Australia\\
    \textsuperscript{\rm 2}Intelligent Medical Computing Laboratory, Faculty of Applied Sciences, Macao Polytechnic University, Macao, China\\
    \textsuperscript{\rm 3}The Chinese University of Hong Kong, Hong Kong SAR, China\\

    \textit{tong.chen1@sydney.edu.au}\\
    \textcolor{magenta}{\textbf{https://github.com/DavisMeee/EndoIR}}

}

\begin{document}

\maketitle

\begin{abstract}
Endoscopic images often suffer from diverse and co-occurring degradations such as low lighting, smoke, and bleeding, which obscure critical clinical details. Existing restoration methods are typically task-specific and often require prior knowledge of the degradation type, limiting their robustness in real-world clinical use. We propose EndoIR, an all-in-one, degradation-agnostic diffusion-based framework that restores multiple degradation types using a single model. EndoIR introduces a Dual-Domain Prompter that extracts joint spatial–frequency features, coupled with an adaptive embedding that encodes both shared and task-specific cues as conditioning for denoising. To mitigate feature confusion in conventional concatenation-based conditioning, we design a Dual-Stream Diffusion architecture that processes clean and degraded inputs separately, with a Rectified Fusion Block integrating them in a structured, degradation-aware manner. Furthermore, Noise-Aware Routing Block improves efficiency by dynamically selecting only noise-relevant features during denoising. Experiments on SegSTRONG-C and CEC datasets demonstrate that EndoIR achieves state-of-the-art performance across multiple degradation scenarios while using fewer parameters than strong baselines, and downstream segmentation experiments confirm its clinical utility.

\end{abstract}

\section{Introduction}

Endoscopic imaging is indispensable for gastrointestinal diagnosis and surgical guidance. However, in real-world clinical settings, captured images are often degraded by multiple co-occurring factors such as low illumination, smoke, and bleeding ~\cite{bai2023llcaps,pan2022desmoke,ramirez2002detection}. These degradations obscure fine anatomical structures, reduce visibility of surgical tools, and impair downstream computational analysis. Restoring such degraded images is therefore essential for improving both the diagnostic quality and the safety of endoscopic procedures.

A variety of restoration approaches have been developed over the years~\cite{garcia2023multi,ma2021structure,guo2024pixel}. Early methods largely targeted single degradations in isolation, such as illumination enhancement, smoke removal, or denoising, typically using task-specific CNNs, GANs, or transformer-based architectures~\cite{wang2024desmoking,pan2022desmoke,su2023multi}. While effective for their intended degradation type, these models cannot handle complex, mixed degradations without retraining or explicit knowledge of the degradation type. More recent research in the broader computer vision community has explored all-in-one restoration frameworks, aiming to address multiple degradations with a single unified model.
\begin{figure}
    \centering
    \includegraphics[width=1\linewidth, trim=0 0 0 0]{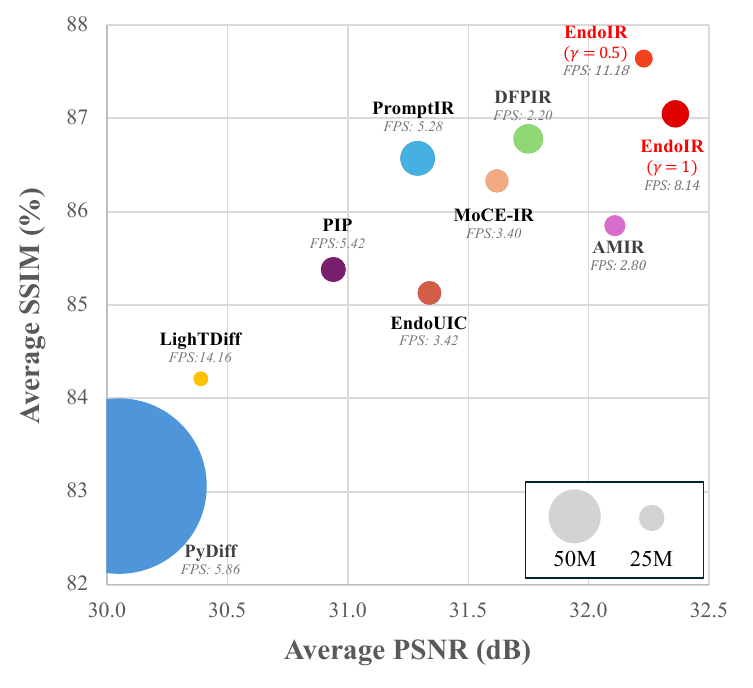}
    \caption{Comparison of average restoration accuracy (PSNR/SSIM), model size, and inference speed (FPS, annotated above each marker). Both our EndoIR ($\gamma=0.5$) and ($\gamma=1$) achieve state‑of‑the‑art restoration quality, while the $\gamma=0.5$ variant offers a compact model size and high inference speed, making it well‑suited for real‑time clinical use.}
    \label{fig:intro}
\end{figure}

Despite these advances, three limitations remain prominent. \underline{First}, many unified methods rely on explicit degradation-type input or accurate prompt extraction. For example, prompt-based methods such as PromptIR~\cite{potlapalli2023promptir} and AutoDIR~\cite{jiang2024autodir} condition restoration on learned visual or language prompts, but these approaches require either prior knowledge of the degradation type or robust degradation classification, which is unreliable in dynamic surgical scenes. \underline{Second}, routing-based~\cite{ren2024moe} architectures such as AMIR [1]te{yang2024all} select restoration paths from degradation-specific cues. While effective for distinguishing degradation types, this can bias the model toward degradation modeling at the expense of preserving fine anatomical structures---a critical concern in medical images where subtle structural fidelity is essential. \underline{Third}, in the case of diffusion-based methods~\cite{chen2024lightdiff,xia2023diffir}, conventional concatenation-based conditioning merges degraded and noisy inputs early in the network. This can confuse the model with mixed feature distributions, particularly when degradations co-occur, leading to unstable optimization and reduced restoration quality. A detailed side-by-side comparison of these approaches and our EndoIR is provided in Supplementary Table~1, highlighting the differences in degradation handling, conditioning strategy, and efficiency.

To address these challenges, we introduce \textbf{EndoIR}, the first all-in-one, degradation-agnostic diffusion framework tailored for complex endoscopic image restoration. EndoIR is designed to operate without any prior knowledge of the degradation type, while effectively disentangling and leveraging both degradation-specific and content-preserving information. Central to our approach is a \textbf{Dual-Domain Prompter} that jointly extracts spatial and frequency features to produce informative restoration prompts, coupled with a \textbf{Task Adaptive Embedding} mechanism that separates shared anatomical content from degradation-specific cues. To overcome feature confusion in conventional conditioning, we design a \textbf{Dual-Stream Diffusion} architecture that processes clean and degraded inputs independently, followed by a \textbf{Rectified Fusion Block} that integrates them in a structured, degradation-aware manner. Finally, to improve computational efficiency, we propose a \textbf{Noise-Aware Routing Block} that dynamically selects only noise-relevant features for denoising, reducing redundant computation while preserving restoration quality.

\vspace{0.3em}
\noindent{Our main contributions are summarized as follows:
\begin{itemize}
    \item We propose degradation-agnostic restoration without labels via a Dual-Domain Prompter and task adaptive embedding for joint spatial-frequency prompt learning.
    \item We introduce a disentangled dual-stream conditioning mechanism for stable diffusion restoration, with a Rectified Fusion Block for structured feature integration.
    \item We design an efficient, noise-aware decoding strategy using the Noise-Aware Routing Block that reduces computation without compromising quality.
    \item We conduct comprehensive validation on challenging endoscopic benchmarks, achieving state-of-the-art performance on SegSTRONG-C and CEC datasets and demonstrating strong generalization in downstream surgical tool segmentation tasks.
\end{itemize}

\section{Related Work}
In general scenarios for all-in-one restoration, early works addressed diverse degradation scenarios using corruption-agnostic frameworks, such as contrastive degradation encoders and degradation-guided decoders~\cite{li2022air}. Subsequent methods like RAM~\cite{qin2024restore} introduced masked image modeling for intrinsic feature extraction, while ADMS~\cite{park2023all} utilized adaptive filters to handle unknown degradations with minimal overhead.
Recent studies explored adaptive prompts to guide restoration. PromptIR~\cite{potlapalli2023promptir} and PIP~\cite{li2023prompt} developed prompt-based modules to dynamically encode degradation information, combining high-level and low-level prompts through dedicated interaction mechanisms.
Vision-language approaches also advanced the field by injecting semantic cues. AutoDIR~\cite{jiang2024autodir} integrated a vision-language module for open-vocabulary degradation recognition with a diffusion model guided by text prompts. MPerceiver~\cite{ai2024multimodal} leveraged CLIP for improved zero-shot and few-shot performance.
To mitigate task interference and data scarcity, DFPIR~\cite{tian2025degradation} introduced feature perturbations guided by degradation prompts, while FoundIR~\cite{li2024foundir} constructed a large real-world dataset and a two-stage diffusion-refinement framework. Defusion~\cite{luo2025visual} further refined diffusion models with degradation-specific visual priors, focusing on visual rather than language cues.

In summary, unified frameworks and more recent prompt- and diffusion-based approaches have advanced all-in-one restoration with strong adaptability and robustness to diverse degradations. However, in medical imaging, degradation types and patterns often differ from those in natural images. Medical images typically have more regular structures and are highly sensitive to small changes, especially at lesion borders and in fine textures. Inaccurate restoration can cause errors in tasks like segmentation and diagnosis. Only a few studies address all-in-one restoration for medical images~\cite{li2018joint}. 
Recently, AMIR~\cite{yang2024all} introduced a task-adaptive routing network for multiple tasks, such as MRI super-resolution, CT denoising, and PET synthesis. 
Subsequent work further explores new paradigms in model architecture design~\cite{yang2025restore,chen2025all}.
Nevertheless, in endoscopic scenarios, most work still targets specific tasks, like low-light enhancement~\cite{bai2023llcaps,chen2024lightdiff}, super-resolution~\cite{chen2022dynamic,liu2023mda}, or desmoking~\cite{pan2022desmoke,wu2024self}, and optimizes each task in isolation. Few methods aim for a unified solution. For example, EndoUIC~\cite{bai2024endouic} provides unified enhancement for illumination only. Yet, degradations such as low light, smoke, and blood often occur together in real endoscopy~\cite{ding2024segstrong}, creating a clear need for an all‑in‑one model that can robustly restore images under multiple degradations for accurate diagnosis and safe clinical decision‑making.

\section{Methodology}

\begin{figure*}[t]
    \centering
    \includegraphics[width=1.\linewidth, trim=0 0 0 0]{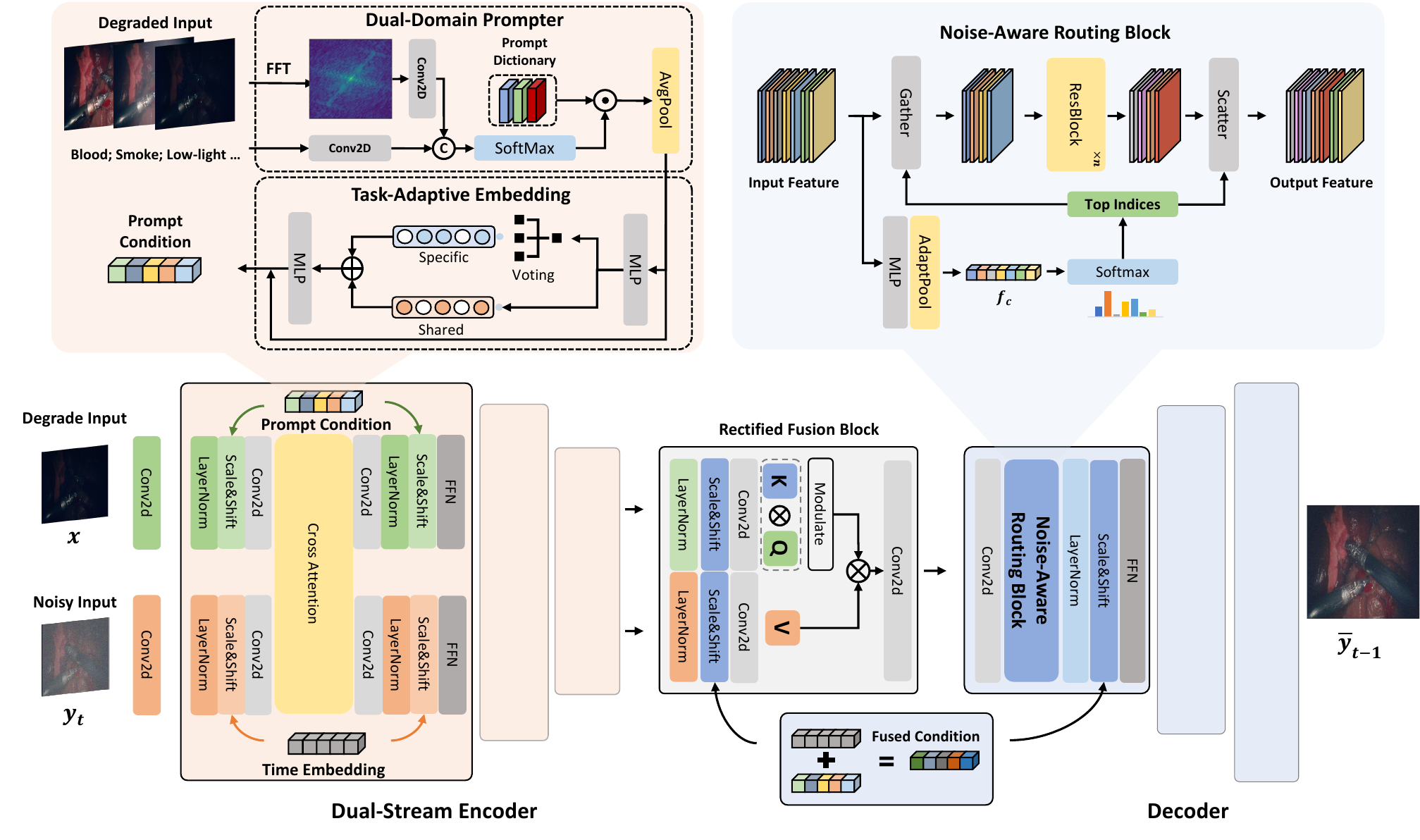}
    \caption{{Overview of our EndoIR framework: Our denoising block consists of a Dual-Domain Prompter(DDP) to generate fine-grained prompt guidance; Task Adaptive Embedding(TAE) to dynamically learn task-specific conditions; Dual-Stream Encoder(DSE) to enable the dual-stream interaction for decoupled feature learning; Rectified Fusion Block(RFB) is designed to fuse dual-stream features; Noise-Aware Routing Block(NARB) for efficient feature refinement.}}
    \label{fig:main}
\end{figure*}

\textbf{Overview.}
The pipeline of our EndoIR framework is presented in Fig.~\ref{fig:main}. Our EndoIR framework is based on the pyramid DDIM model~\cite{zhou2023pyramid}, which integrates a diffusion method to perform sampling in a pyramid resolution style to achieve faster sampling. Its processes can be expressed as follows:
\begin{equation}
\begin{aligned}
&q(y_{t} \mid y_{t-1}) := \mathcal{N}\left(y_{t}; \sqrt{\hat{\alpha_{t}}}(y_t\downarrow r_{t-1}/r_{t}), \beta_{t}\textbf{I}\right), \\
\end{aligned}
\end{equation}
\begin{equation}
\begin{aligned}
&p(y_{t-1} \mid y_t) := \mathcal{N}(y_{t-1};\sqrt{\hat{\alpha}_{t-1}} f_\theta(y_t \uparrow r_{t}/r_{t-1}) \\&+ \frac{1 - \hat{\alpha}_{t-1}}{1 - \hat{\alpha}_t} y_t,
\frac{\beta_t(1 - \hat{\alpha}_{t-1})}{1 - \hat{\alpha}_t}\, \mathbf{I}
\Big),
\end{aligned}
\end{equation}
where $y_t$ represents the image at time step $t$, $\beta_t$ is the noise variance, and $\hat{\alpha}_t := \prod^{t}_{i=1} (1-\beta_i)$ is the cumulative product of noise schedules given by $\beta_t$. The denoising framework of our diffusion model adopts a Dual-Unet~\cite{ronneberger2015u} architecture, consisting of our Dual-Stream Encoder and a Decoder with Noise-Aware Routing Blocks. The Rectified Fusion Block shall perform the fusion of dual-stream intermediate features. Additionally, the Dual-Domain Prompter and Task Adaptive Embedding are designed to achieve auxiliary feature representation for precision denoising guidance.

\noindent \textbf{Dual-Domain Prompter.} Prompt-based approaches often rely solely on spatial features, missing the complementary frequency cues that explicitly capture global structural integrity. Frequency features, obtained via Fast Fourier Transform (FFT), reveal global spectral signatures of degradation — for example, blood introduces localized low frequency dominance due to large homogeneous regions, smoke attenuates mid and high frequency components by veiling fine details, and low light shifts the overall spectrum. By fusing these complementary cues, we propose Dual-Domain Prompter to form richer degradation-aware prompts that encode both the spatial and the spectral characteristics of corruption. These prompts guide the restoration network to adapt effectively to diverse degradations without requiring explicit degradation labels. The effectiveness of this spatial frequency fusion is validated in our ablation results~\ref{fig:t-SNE}. First, we initialize a set of learnable parameters as the prompt dictionary $\mathbb{D}$. Then, we extract joint representations by integrating information from the spatial and frequency domains. After combining these representations with $\mathbb{D}$, we apply average pooling to obtain the prompt output. The Dual-Domain Prompter can be formulated as:
\begin{equation}
\begin{aligned}
&F_{img} = Conv2D(x) \\
&F_{freq} = Conv2D(\mathbf{FFT}(x))\\
&\mathbb{P} = \text{AvgPool}\{\text{Softmax}[F_{img} + F_{freq}] \cdot \mathbb{D} \}
\end{aligned}
\end{equation}
in which $x$ represent the input image and $\mathbb{P}$ is the output prompt parameter. The output prompt $\mathbb{P}$ is further propagated through the Task Adaptive Embedding to generate the conditions for controlling the diffusion process.

\noindent \textbf{Task Adaptive Embedding.} In Surgical imaging scenarios, the clean images often exhibit relatively consistent contents. When using a unified routing scheme, there is a risk that the model may overfit to degradation-specific cues, potentially leading to semantic drift or incorrect restoration. To mitigate this, we propose Task-Adaptive Embedding (TAE), which balances general content preservation with corruption-specific adaptation.

The task adaptive embedding module receives the DDP output $\mathbb{P}$ as input and processes it through two parallel branches. The shared branch captures global, content-preserving semantics that remain stable across different types of degradation. The specific branch consists of multiple modules, each specialized in handling a particular corruption pattern. A soft voting mechanism selects the most degraded-relevant branch based on the prompt itself. Formally, the task-aware embedding is given by:

\begin{equation}
    E_{\text{task}} = \sum_{i=1}^{n}MLP_{i}(\mathbb{P}) + \sum_{k=1}^{K}MLP_{k}(\mathbb{P}) \cdot Top\mathbf{K}(\text{Softmax}(\mathbb{P}))
\end{equation}
This disentangled embedding structure ensures that the model adapts flexibly to varying corruption types while maintaining fidelity to anatomical content. This hybrid embedding scheme enhances the model’s ability for generate precise, condition-aware representations for guiding the denoising process within diffusion model.

\noindent \textbf{Dual-Stream Encoder.}  
Traditional conditional diffusion models often adopt a simple concatenation of the corrupted image as guidance, implicitly assuming that both follow similar distributions. However, since various corrupted inputs lie in distinct distribution domains, naïve concatenation can disrupt the intrinsic distribution of clean images, thereby introducing uncertainty and compromising the effectiveness of the denoising process. To address this, we propose a Dual-Stream Encoder(DSE) that separately encodes the corrupted image and the noise-added image, thereby disentangling degradation-specific information and facilitating more accurate guidance for restoration.

As illustrated in Fig.~\ref{fig:main}(d), given a corrupted input image $x$ and its corresponding noise-added counterpart $y_t$, we first embed both images into feature representations using separate convolution layers, followed by Layer Normalization:
\begin{equation}
\begin{aligned}
    F_n' &= \text{Conv2D}(\text{LN}(\text{Conv2D}(n))), \quad n \in \{x, y\}
\end{aligned}
\end{equation}

These two features are then jointly processed via a cross-attention module. Specifically, the feature maps are concatenated along the channel dimension and treated as the query, key, and value in a self-attention block. Since the attention computation is spatially self-referential, the two different feature streams do not directly cross-interfere. The output is split back into two streams using a chunk operation, which are then passed through feed-forward convolution modules with residual connections to form the final encoded features.

The proposed architecture enables the model to capture both degradation-aware and content-aware representations in a disentangled manner. This design not only enhances the model’s ability to characterise degradation patterns but also promotes the preservation of fine structures and textures during restoration.

\noindent \textbf{Rectified Fusion Block.}
After the dual-stream encoder, both features from the degraded domain and the diffusion domain are supposed to be sufficiently extracted. To effectively integrate these two representations, we introduce the Rectified Fusion Block (RFB), which employs a modulated cross-attention. Specifically, we use the degraded-domain feature $F_{\overline{x}}$ as the source for query and key, and the diffusion-domain feature $F_{\overline{y}}$ as the value:
\begin{equation}
    F_{\overline{n}}' = \text{Conv2D}(\text{LayerNorm}(F_{\overline{n}})), \quad n \in \{x, y\},
\end{equation}
where the output of $F_{\overline{x}}'$ is expanded to twice the channel size of $F_{\overline{y}}'$, allowing splitting into $Q$ and $K$. The chunk function is applied to obtain:
\[
Q, K = \text{chunk}(F_{\overline{x}}'), \quad V = F_{\overline{y}}'
\]

To reduce the domain discrepancy and guide the attention map computed from the degraded domain towards the structure of the diffusion domain, we apply a modulation on the similarity scores:
\begin{equation}
    \text{Attn} = \left[ w_1 \cdot \text{Softmax}(QK^\top) + w_2 \cdot \text{GeLU}(QK^\top) \right] \cdot V,
\end{equation}
where $w_1$ and $w_2$ are learnable scalars that balance sharp and smooth attention responses, aligning the degraded-domain attention with the structure of the diffusion domain. Finally, the output is processed through a residual convolution and a feed-forward network to yield the fused feature:
\begin{equation}
    F_f = \text{FFN} \left( \text{Conv2D}(\text{Attn}) + F_{\overline{x}} \right).
\end{equation}

This fusion block rectifies the features of both domains, producing a unified feature that captures both the degradation context and structural priors.

\noindent \textbf{Noise-Aware Routing Block.}
Standard decoders process all feature channels at every denoising step, even though degraded and clean images share substantial structural similarity and many channels may contain features already close to clean. This wastes computation and, in multi‑degradation restoration, allows degradation‑irrelevant features to interfere with reconstruction. 
\begin{algorithm}[t]
\caption{\textbf{Noise-Aware Routing Block}}
\label{alg:narb}
\hspace*{0.02in}{\textbf{Input}}: $F_{\text{in}} \in \mathbb{R}^{C \times H \times W}$;\\
Selection ratio $\gamma \in (0, 1]$;\\
Number of ResBlocks $N_{\text{res}}$ \\
\hspace*{0.02in}{\textbf{Output}}: Refined feature $F_{\text{out}} \in \mathbb{R}^{C \times H \times W}$ \\
\begin{algorithmic}[1]
\STATE $f_c \gets \text{AdaptiveAvgPool}(F_{\text{in}})$
\STATE $a \gets \text{Softmax}(\text{Linear}(\text{Linear}(f_c))) \in \mathbb{R}^C$ \\
\STATE //     Top-$k$ Selection
\STATE $\mathcal{I}_{\text{select}} \gets \text{Top}k(a, \gamma \cdot C)$ 
\STATE $F_{\text{select}} \gets \text{Gather}(F_{\text{in}}, \mathcal{I}_{\text{select}})$ \\
\STATE //    Feature Refinement

\FOR{$n$ in Number of ResBlocks $N_{\text{res}}$}
\STATE $F'_{\text{select}} \gets \text{ResBlock}_n(F_{\text{select}})$
\ENDFOR
\STATE $F_{\text{out}} \gets \text{Scatter}(F'_{\text{select}}, \mathcal{I}_{\text{select}}, F_{\text{in}})$
\RETURN {$F_{\text{out}}$}
\end{algorithmic}
\end{algorithm}

\begin{table*}[t]
	\caption{
Single-mode image restoration (blood removal, low-light image enhancement, and desmoking) and all-in-one image restoration performance comparison of our EndoIR against SOTA restoration methods on the SegSTRONG-C dataset. The all-in-one restoration methods are trained only once, while the original single-task methods are trained separately for each task.
	} 
 	\centering
    \renewcommand{\arraystretch}{1.2}
	\label{tab:segstrong}  
\resizebox{\textwidth}{!}{	
\begin{tabular}{c|cc|ccc|ccc|ccc|ccc}
\hline

\multirow{2}{*}{\textbf{Models}} &\multirow{2}{*}{\textbf{Params(M)}} &\multirow{2}{*}{\textbf{FPS}}& \multicolumn{3}{c|}{Blood}          & \multicolumn{3}{c|}{Low-Light} & \multicolumn{3}{c|}{Smoke} & \multicolumn{3}{c}{Average}  \\ \cline{4-15} & & &PSNR $\uparrow$ & SSIM $\uparrow$  & LPIPS $\downarrow$ & PSNR $\uparrow$ & SSIM $\uparrow$  & LPIPS $\downarrow$ & PSNR $\uparrow$ & SSIM $\uparrow$  & LPIPS $\downarrow$&PSNR $\uparrow$ & SSIM $\uparrow$  & LPIPS $\downarrow$ 
\\ \hline 
PyDiff~\cite{zhou2023pyramid}          &97.9 &5.86   &29.77 & 82.80 & 0.1309 &30.72&	85.63&0.0993 
 & 30.36 & 83.10& 0.1406 & 30.05 & 83.06 &0.1296 \\
LighTDiff~\cite{chen2024lightdiff} &\underline{18.6} &\textbf{14.16} &30.75	&84.93	&0.0858	&31.79	&86.47	&0.07526	&30.71	&84.79	&0.0773	&30.39	&84.21	&0.1011\\
EndoUIC~\cite{bai2024endouic}     &25.8  &3.42    &31.15 &84.19 & 0.0763 & 32.48 & 86.22 & 0.0644 & 31.01 & 85.98 & 0.0616 &31.34	&85.13	&0.0694 \\
LLCaps~\cite{bai2023llcaps}     &120.0  &2.40  &30.46	&84.13	&0.1280	&31.00	&85.17	&0.1530	&31.60	&86.37	&0.1128	&31.26 &85.62	&0.1268  \\
Diff-LOL~\cite{jiang2023low}    &22.1  &2.29  &30.38	&82.24	&0.1024	&28.90	&80.16	&0.1790&	23.06	&72.43	&0.1647	&27.41	&78.57	&0.1496\\
LA-Net~\cite{yang2023learning}   &\textbf{0.57}  &2.05    &28.14 	&81.26 	&0.1549 &27.15 	&79.68 	&0.1974 	&29.06 	&82.57 	&0.1536 
	&26.81	&78.80	&0.1943  \\ \hline
PromptIR~\cite{potlapalli2023promptir}      &33.0  &5.28   &29.72 & 84.96  & 0.0992 & 32.27 & 87.13 & 0.0741 & 31.88 & 87.61 & 0.0615 & 31.29  & 86.57  & 0.0783 \\
PIP~\cite{li2023prompt}           &26.8   &5.42   &30.02 & 83.37 & 0.0905 & 31.32 & 86.60 & 0.0782 & 31.48 & 86.19 & 0.0646 & 30.94 & 85.38& 0.0778 \\

AMIR~\cite{yang2024all}     &23.54  &2.80  &31.02	&83.53 	&0.0795 &\underline{32.71} &87.19 	&0.0868 	&32.09 	&86.84 	&0.0669 
	&32.11	&85.85	&0.0777  \\
MoCE-IR~\cite{zamfir2025complexity}    &25.35  &3.40   &30.31	&84.78 	&0.1797 &32.19 &86.96 	&0.2054 	&32.37 	&87.26 	&0.1509 &31.62	&86.33	&0.1786  \\
DFPIR~\cite{tian2025degradation}     &29.63  &2.20  &30.57	&85.01 	& 0.0932 & 32.38 & \underline{87.94} 	&0.0768 	&32.32 	&87.40 	&0.0615 &31.75	&86.78	&0.0771  \\  \hline
\textbf{EndoIR ($\gamma=0.5$)} &21.3 &\underline{11.18} & \underline{31.33} & \textbf{86.76} & \underline{0.0759} & 32.64 & \textbf{87.96} & \textbf{0.0679} & \underline{32.71} & \textbf{88.20} & \underline{0.0554} & \underline{32.23} & \textbf{87.64} & \underline{0.0664} \\
% \textbf{EndoIR (75\%)}&22.8 &  &  &  &  &  &  &  &  &  & & &  \\
\textbf{EndoIR ($\gamma=1$)}&28.2 &8.14  &\textbf{31.53} & 85.96 & \textbf{0.0640} & \textbf{32.82} & 87.67 & \underline{0.0693} & \textbf{32.74} & 87.51 & \textbf{0.0551} & \textbf{32.36} & 87.05 & \textbf{0.0628 }\\ \hline
\end{tabular}}
\end{table*}
We propose a Noise‑Aware Routing Block (NARB) that dynamically selects only the top‑$k$ most relevant channels at each step, conditioned on the current noise embedding~\ref{alg:narb}. Given a feature map $F \in \mathbb{R}^{C \times H \times W}$, NARB first computes a compact channel descriptor $f_c$ via global average pooling. This descriptor is passed through a noise‑conditioned gating network to estimate per‑channel relevance $a$. Given the selection ratio $\gamma$, we retain only the most relevant channels $F_{\text{select}} \in \mathbb{R}^{\gamma \cdot C \times H \times W}$, indexed by $\mathcal{I}_{\text{select}}$, where $\mathcal{I}_{\text{select}}$ is determined by selecting the Top-$k$ channels with the highest relevance scores, with $k := \gamma \cdot C$. These selected channels are refined by residual blocks to produce $F'_{\text{select}}$, which are then reinserted into their original positions to form the output feature $F_{\text{out}}$.

Through selective channel refinement, NARB dynamically allocates denoising capacity to the most noise‑affected dimensions, improving efficiency while preserving clean semantic features. This targeted strategy maintains overall restoration fidelity with reduced computation.

\section{Experiments}

\subsection{Experiment Settings}

\noindent \textbf{Datasets.} 

\textbf{All-in-one Restoration:} SegSTRONG-C dataset~\cite{ding2024segstrong} was initially developed for robust segmentation tasks under non-adversarial corruptions. The validation and test sets contain images with three categories of corrupt scenarios: blood, low-brightness, and smoke, in which each category comprises 900 images. We merge and divide these images in 3:1 ratio, resulting in 1350 images per category for the training set and 450 images per category for the test set.

\textbf{Unified Illumination Correction:} CEC dataset~\cite{bai2024endouic} is a capsule endoscopy illumination restoration dataset consisting of underexposed and overexposed images. The annotations were generated by professional photographers using the Adobe SDK. The dataset includes 800 training images and 200 testing images.

\noindent \textbf{Implementation Details.}
We compare our method against SOTA all-in-one and specialized methods, as shown in Table~\ref{tab:segstrong}. The all-in-one restoration methods are trained only \textbf{once}, while the single-task methods are trained \textbf{separately} for each task. The experiments are performed on NVIDIA A40 GPUs. Our model is trained for 100 epochs using Adam with a learning rate of \(2 \times 10^{-4}\) and a batch size of $8$.
% and a decay factor (\(\gamma\)) of 0.3, controlled by the MultiStepLR scheduler. And batch size on per GPU is 8. 
The evaluation metrics include PSNR, SSIM, and LPIPS. Moreover, to address how well the task-adaptive embedding enhances feature discrimination across different restoration tasks, we involve Wilks' Lambda, a statistical metric that quantifies class separability. It can be demonstrated as a significant difference when the value is lower than 0.05. 

We further explore the clinical applicability by performing downstream surgical tool segmentation on the enhanced images of the SegSTRONG-C dataset~\cite{ding2024segstrong}. 
The segmentation model is based on a pre-trained ResNet-101~\cite{he2016deep} backbone and DeeplabV3+~\cite{chen2018deeplabv3plus} decoder, trained for 200 epochs on the SegSTRONG-C training set, and evaluated on the test set. We assess the segmentation results using mean Intersection over Union (mIoU), Dice score (Dice), and Accuracy (ACC).

\begin{table*}[t]
    \caption{Performance comparison across different models with evaluation metrics on CEC Dataset~\cite{bai2024endouic}.}
    \resizebox{\textwidth}{!}{	
    \centering
    \setlength{\tabcolsep}{2.5pt} 
    \renewcommand{\arraystretch}{1} 
    \begin{tabular}{l|*{13}{c}} \hline
    
        \textbf{Models}&

        \makecell[c]{PromptIR}%\\~\cite{potlapalli2023promptir}} 
        &\makecell[c]{PIP}%\\~\cite{li2023prompt} }& 
        &\makecell[c]{PyDiff}%\\~\cite{zhou2023pyramid}} & 
        &\makecell[c]{LighTDiff}%\\~\cite{chen2024lightdiff}} & 
        &\makecell[c]{EndoUIC}%\\~\cite{bai2024endouic} }& 
        &\makecell[c]{LLCaps}%\\~\cite{bai2023llcaps}} & 
        &\makecell[c]{Diff-LOL}%\\~\cite{jiang2023low}} & 
        &\makecell[c]{LA-Net}%\\~\cite{yang2023learning}} & 
        &\makecell[c]{AMIR}%\\~\cite{yang2024amir}} & 
        &\makecell[c]{MoCE-IR}%\\~\cite{yang2024amir} }& 
        &\makecell[c]{DFPIR}%\\~\cite{yang2024amir} }&
        &\makecell[c]{\textbf{EndoIR}\\($\gamma=0.5$)}
        &\makecell[c]{\textbf{EndoIR}\\($\gamma=1$)}  \\  \hline  
        \textbf{PSNR}(dB) $\uparrow$ & 
        28.27 & 25.01 & 30.54 & 29.23 & 26.39 & 27.55 & 28.07 & 15.86 & 32.27 & 29.44 &26.34 & \textbf{32.65} & 32.61 \\
        \textbf{SSIM}(\%) $\uparrow$ & 
        83.14 & 70.09 & 77.07 & 96.14 & 95.64 & 85.95 & 96.13 & 66.40 & 96.43 & 93.27 &84.17 & 96.98 & \textbf{97.48} \\
        \textbf{LPIPS} $\downarrow$ & 
    0.0717 & 0.1833 & 0.0796 & 0.0752 & 0.1079 & 0.2366 & 0.0883 & 0.4233 & 0.0512 & 0.1257 &0.0795 & 0.0481 & \textbf{0.0451} \\ \hline    \end{tabular}}
    \label{tab:cec_table}
\end{table*}
\begin{table*}[t]
    \caption{Downstream task performance across different models on SegSTRONG-C Dataset~\cite{ding2024segstrong}.}
    \resizebox{\textwidth}{!}{	
    \centering
    \setlength{\tabcolsep}{2.5pt} 
    \renewcommand{\arraystretch}{1} 
    \begin{tabular}{l|*{13}{c}} \hline

        \textbf{Models}&

        \makecell[c]{PromptIR}%\\~\cite{potlapalli2023promptir}} 
        &\makecell[c]{PIP}%\\~\cite{li2023prompt} }& 
        &\makecell[c]{PyDiff}%\\~\cite{zhou2023pyramid}} & 
        &\makecell[c]{LighTDiff}%\\~\cite{chen2024lightdiff}} & 
        &\makecell[c]{EndoUIC}%\\~\cite{bai2024endouic} }& 
        &\makecell[c]{LLCaps}%\\~\cite{bai2023llcaps}} & 
        &\makecell[c]{Diff-LOL}%\\~\cite{jiang2023low}} & 
        &\makecell[c]{LA-Net}%\\~\cite{yang2023learning}} & 
        &\makecell[c]{AMIR}%\\~\cite{yang2024amir}} & 
        &\makecell[c]{MoCE-IR}%\\~\cite{yang2024amir} }& 
        &\makecell[c]{DFPIR}%\\~\cite{yang2024amir} }&
        &\makecell[c]{\textbf{EndoIR}\\($\gamma=0.5$)}
        &\makecell[c]{\textbf{EndoIR}\\($\gamma=1$)}  \\  \hline  % 
        \textbf{Dice} $\uparrow$ & 
        95.09 & 94.97 & 95.53 & 95.03 & 91.38 & 92.63 & 92.59 & 91.31 & 95.13 & 95.35 & 95.47 &  95.54 & \textbf{96.22} \\
        \textbf{mIoU} $\uparrow$ & 
        90.41 &  90.56 & 91.79 & 90.43 & 84.29 & 84.20 & 84.96 & 84.19 & 90.78 & 90.80 & 91.39 & \textbf{91.51}   & 91.40 \\
        \textbf{ACC} $\uparrow$ & 
        98.15 & 98.06 & 98.75 & 98.45 & 96.11 & 96.43 & 96.25 & 95.40 & 98.16 & 98.18 & 98.27 & 98.31 & \textbf{98.97} \\ \hline
    \end{tabular}}
    \label{tab:seg_table}
\end{table*}

\begin{figure*}[!t]
    \centering
    \setlength{\abovecaptionskip}{0.8cm}
    \includegraphics[width=1\linewidth, trim=0 30 0 10]{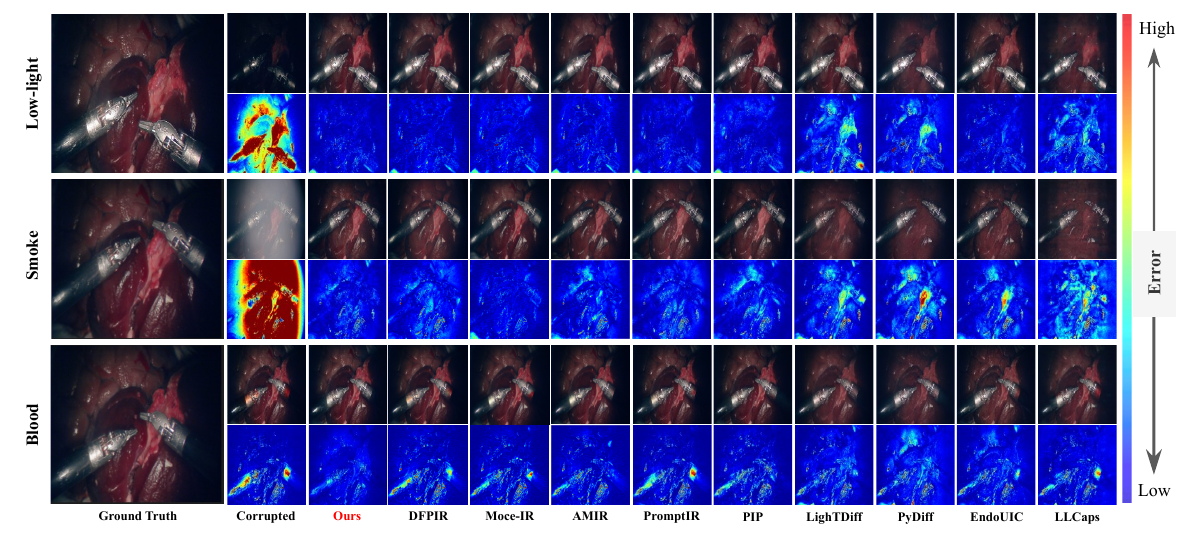}
    \caption{The quantitative visualizations and error maps on the SegSTRONG-C~\cite{ding2024segstrong} dataset. Blue and red represent low and high error, respectively. (Zoom in to see details.)
    }
    \label{fig:IMG_visual}
\end{figure*}

\subsection{Experimental Results}
\label{sec:results}

\noindent \textbf{Restoration Results.}
Table~\ref{tab:segstrong} quantitatively compares the restoration performance of our EndoIR framework against a range of SOTA methods across the unified all-in-one restoration setting on the SegSTRONG-C dataset. Our EndoIR consistently outperforms prior advanced restoration models across all evaluation metrics. Notably, even compared to the strong medical restoration baseline AMIR~\cite{yang2023learning}, which exhibits competitive results in all subtasks, indicating that our method produces more perceptually faithful reconstructions.
Specifically, EndoIR achieves a peak PSNR of 32.82 dB and SSIM of 87.67\% on low-light enhancement, slightly surpassing AMIR and clearly outperforming illumination restoration approaches such as EndoUIC~\cite{bai2024endouic} and LLCaps~\cite{bai2023llcaps}. In the smoke and blood removal task, EndoIR also attains excellent structural preservation and perceptual similarity, showcasing the model's robustness in handling various degradation sources.
Besides, EndoIR delivers the highest average performance, confirming its capability in capturing task-shared and task-specific features simultaneously. Our method demonstrates a clear edge in preserving fine details and perceptual quality under compound degradation scenarios. Figure~\ref{fig:IMG_visual} further supports our quantitative findings, where EndoIR exhibits the bluest error maps among all baselines, indicating the least pixel-wise deviation from the ground truth. We also evaluate EndoIR on a unified illumination correction benchmark (Table~\ref{tab:cec_table}), where it achieves a significant PSNR improvement of 1.96 dB and a 0.68\% SSIM gain over the SOTA, reaffirming the generalizability of our method across different datasets.
These comprehensive results demonstrate the strong restoration capacity of EndoIR and highlight its potential for robust deployment in real clinical endoscopic applications.

\noindent \textbf{Downstream Evaluation.} Furthermore, downstream segmentation tasks are conducted to assess the effectiveness of EndoIR in joint medical image restoration. As presented in Table~\ref{tab:seg_table}, EndoIR surpasses competing methodologies in surgical instrument segmentation, demonstrating its exceptional ability to preserve instrumental structures. It consistently outperforms all SOTA methods in mIoU and Dice scores (91.40/96.22), thereby emphasizing its advanced abilities in restoring images and preserving critical details.

\noindent \textbf{Efficiency.} Compared with other SOTA methods, our model also demonstrates strong efficiency in addition to the high restoration performance. As shown in Table~\ref{tab:segstrong}, the FPS of our model is second only to LightDiff. Table~\ref{tab:segstrong} also demonstrates that EndoIR can achieve strong restoration performance with a relatively lower number of parameters, highlighting its excellent potential for real-world deployment.

\subsection{Ablation Study} 

\begin{table}[!h]
\centering
\caption{Ablation study on the Dual-Domain Prompter with employing frequency embedding $F_{freq}$ and image embedding $F_{img}$ on the SegSTRONG-C dataset.}
\label{tab:ablation_prompt}
\setlength{\tabcolsep}{9pt} 
\begin{tabular}{c|c|ccc}
\hline
\makecell{$F_{freq}$} & \makecell{$F_{img}$} & PSNR $\uparrow$ & SSIM $\uparrow$ & LPIPS $\downarrow$ \\ \hline
$\boldsymbol{\times}$ & $\boldsymbol{\times}$ & 31.19 & 85.98 & 0.0712 \\
$\boldsymbol{\times}$ & $\checkmark$ & 32.20 & 86.44 & 0.0709 \\
$\checkmark$ & $\boldsymbol{\times}$ & 32.11 & 86.85 & 0.0675 \\
$\checkmark$ & $\checkmark$ & \textbf{32.36} & \textbf{87.05} & \textbf{0.0628} \\ \hline
\end{tabular}
\end{table}

\begin{table}[t]
\centering
\caption{Ablation study on SegSTRONG-C~\cite{ding2024segstrong}. We remove (i) the Task Adaptive Embedding, (ii) the Dual-Stream Encoder, (iii) and the Rectified Fusion Block.}
\label{tab:ablation_module}
\renewcommand{\arraystretch}{1.1} 
\setlength{\tabcolsep}{7pt} 
\begin{tabular}{c|c|c|ccc c} \hline
TAE & DSE & RFB & PSNR $\uparrow$ & SSIM $\uparrow$ & LPIPS $\downarrow$ \\ \hline
$\boldsymbol{\times}$ & $\boldsymbol{\times}$ & $\boldsymbol{\times}$ & 30.39 & 84.21 & 0.1011 \\
$\checkmark$ & $\boldsymbol{\times}$ & $\boldsymbol{\times}$ & 31.40 & 85.09 & 0.0741 \\
$\boldsymbol{\times}$ & $\checkmark$ & $\boldsymbol{\times}$ & 31.96 & 86.83 & 0.0703 \\
$\boldsymbol{\times}$ & $\boldsymbol{\times}$ & $\checkmark$ & 31.56 & 86.03 & 0.0764 \\
$\checkmark$ & $\checkmark$ & $\boldsymbol{\times}$ & 32.20 & 86.94 & 0.0649 \\
$\checkmark$ & $\boldsymbol{\times}$ & $\checkmark$ & 31.51 & 86.22 & 0.0706 \\
$\boldsymbol{\times}$ & $\checkmark$ & $\checkmark$ & 32.10 & 86.62 & 0.0654 \\
$\checkmark$ & $\checkmark$ & $\checkmark$ & \textbf{32.36} & \textbf{87.05} & \textbf{0.0628} \\ \hline
\end{tabular}
\end{table}

\noindent \textbf{Components Ablation.} Our ablation study in Table~\ref{tab:ablation_module} further shows improvements introduced by our proposed modules. After involving the Dual-Stream Encoder, the performance greatly improved, while the Recified Fusion Block for fine-grained fuses the feature to further improve restoration quality. Our Task Adaptive Embedding provides task-related information for the frame and assists the final pipeline in gaining the best performance among all the evaluation metrics. Table~\ref{tab:ablation_prompt} verifies the prompt effects and shows that the best performance occurs when both spatial and frequency domains are involved.

\noindent \textbf{Frequency Transformation.} We further conduct experiments to compare the effectiveness of different frequency transformations (FFT and Wavelet Transformation) for Dual-Domain Prompter. As shown in Table~\ref{tab:ablation_frequency}, FFT demonstrates superior performance, outperforming the Wavelet Transform by 0.40 dB in PSNR, 0.88\% in SSIM, and 0.0048 in LPIPS. We attribute this improvement to the ability of FFT to capture global frequency patterns and provide more comprehensive frequency domain information, which is more beneficial compared to wavelet transformation. 

\begin{table}[h]
\centering
\caption{Ablation study on the Dual-Domain Prompter by employing different frequency transformation methods on the SegSTRONG-C dataset.}
\label{tab:ablation_frequency}
\renewcommand{\arraystretch}{1.1}
\setlength{\tabcolsep}{9pt} 
\begin{tabular}{c|c|ccc} \hline
\makecell{FFT} & \makecell{Wavelet} & PSNR $\uparrow$ & SSIM $\uparrow$ & LPIPS $\downarrow$ \\ \hline
$\boldsymbol{-}$ & $\checkmark$ & 31.96 & 86.17 & 0.0676 \\

$\checkmark$ & $\boldsymbol{-}$ & \textbf{32.36} & \textbf{87.05} & \textbf{0.0628} \\ \hline
\end{tabular}
\end{table}

\noindent \textbf{Statistical Significance.} 
Figure~\ref{fig:t-SNE} shows the t-SNE visualization of images from different degradation types, before and after the Task Adaptive Embedding. It can be seen that the Task Adaptive Embedding clearly separates the features of the three types of degradations. In addition, Wilks' Lambda analysis~\cite{rao1951asymptotic} reveals a significant difference of 40 times between inter-task discriminability ($\lambda$ = 0.0170) and intra-task variability ($\lambda$ = 0.6907). This large statistical separation shows that the embedding mechanism can preserve important task-specific information while suppressing task-irrelevant features. This ability is especially important in multi-task learning, where representations need to keep different task identities but still support knowledge sharing. Such controlled separation of features helps improve model generalization and reduces interference between tasks.

\begin{figure}[h]
    \centering
    \includegraphics[width=0.9\linewidth, trim=0 100 0 80]{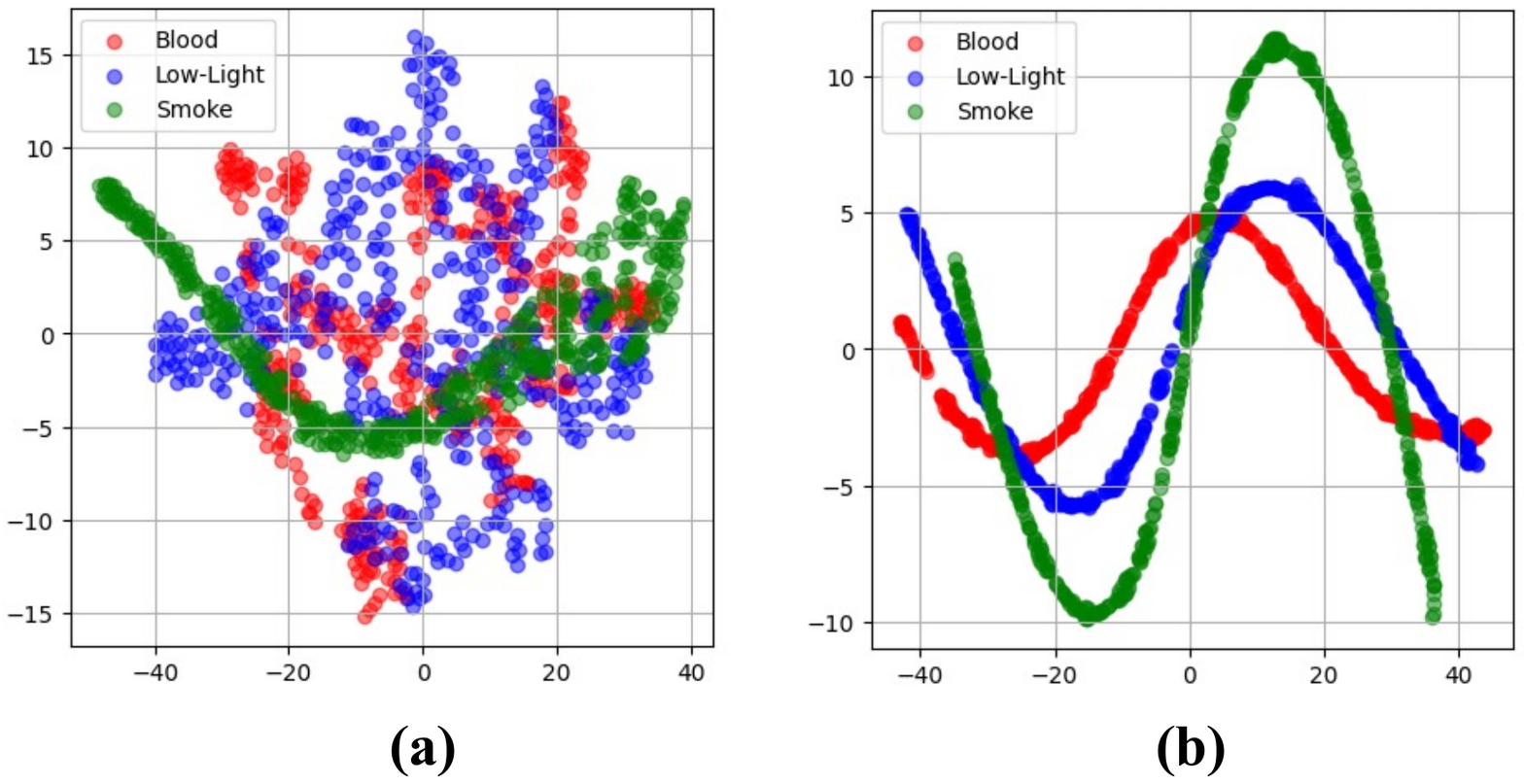}
    \caption{{(a) and (b) shows t-SNE visualization between different tasks: (a) Output feature after passing Dual-Domain Prompter; (b) Task Adaptive Embedding output condition.} }%  (c) demonstrates model performance under different Feature Ratio $\gamma$.}
    \label{fig:t-SNE}
\end{figure}

\noindent \textbf{Feature Selection Ratio.} As shown in Fig.~\ref{fig:plot}, we empirically chose 50\% as the Top-$k$ selection ratio to achieve better performance. Meanwhile, when 100\% features are used, the performance drops significantly. This shows that not all features are noise-relevant, and using all of them leads to redundancy. It is better to select only a representative subset.

\begin{figure}[h]
    \centering
    \includegraphics[width=0.8\linewidth, trim=0 20 0 0]{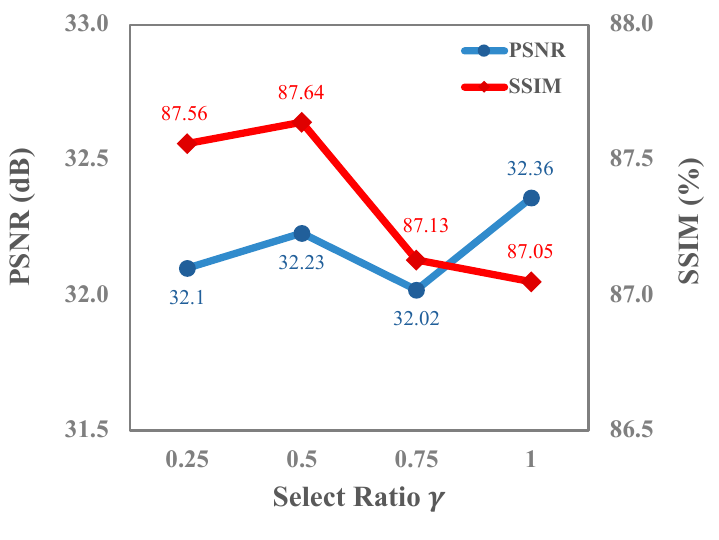}
    \caption{{Ablation study on Noise-Aware Routing Block under different ratio $\gamma$ on the SegSTRONG-C dataset.}}
    \label{fig:plot}
\end{figure}
\section{Conclusion}
\label{sec:conclusion}
In this work, we propose EndoIR, an all-in-one, degradation-agnostic diffusion framework for endoscopic image restoration. EndoIR integrates a Dual-Domain Prompter and task-adaptive embedding to provide robust conditioning, while a Dual-Stream Encoder and Rectified Fusion Block disentangle and integrate degraded and clean representations. Noise-Aware Routing Block further improves efficiency by selectively refining noise-relevant channels. EndoIR sets new benchmarks in endoscopic image restoration, with further tests in segmentation tasks underscoring its potential to enhance surgical precision.
EndoIR shows promising performance across individual degradation types and frame-level restoration, while real-world endoscopic scenes often involve compound degradations and require temporal coherence for reliable video analysis. Future efforts will aim to expand its capabilities to address multi-source compound degradations and incorporate temporal consistency for endoscopic video restoration.

\bibliography{reference}{}

@article{ding2024segstrong,
  title={SegSTRONG-C: Segmenting Surgical Tools Robustly On Non-adversarial Generated Corruptions--An EndoVis' 24 Challenge},
  author={Ding, Hao and Lu, Tuxun and Zhang, Yuqian and Liang, Ruixing and Shu, Hongchao and Seenivasan, Lalithkumar and Long, Yonghao and Dou, Qi and Gao, Cong and Unberath, Mathias},
  journal={arXiv preprint arXiv:2407.11906},
  year={2024}
}

@ARTICLE{guo2024pixel,
  author={Guo, Guangyu and Zhang, Dingwen and Han, Longfei and Liu, Nian and Cheng, Ming-Ming and Han, Junwei},
  journal={IEEE Transactions on Pattern Analysis and Machine Intelligence}, 
  title={Pixel Distillation: Cost-Flexible Distillation Across Image Sizes and Heterogeneous Networks}, 
  year={2024},
  volume={46},
  number={12},
  pages={9536-9550},
  doi={10.1109/TPAMI.2024.3421277}}

@inproceedings{ronneberger2015u,
  title={U-net: Convolutional networks for biomedical image segmentation},
  author={Ronneberger, Olaf and Fischer, Philipp and Brox, Thomas},
  booktitle={Medical Image Computing and Computer-Assisted Intervention--MICCAI 2015: 18th International Conference, Munich, Germany, October 5-9, 2015, Proceedings, Part III 18},
  pages={234--241},
  year={2015},
  organization={Springer}
}

@inproceedings{he2016deep,
  title={Deep residual learning for image recognition},
  author={He, Kaiming and Zhang, Xiangyu and Ren, Shaoqing and Sun, Jian},
  booktitle={Proceedings of the IEEE conference on computer vision and pattern recognition},
  pages={770},
  year={2016}
}

@article{chen2022dynamic,
  title={Dynamic Depth-Aware Network for Endoscopy Super-Resolution},
  author={Chen, Wenting and Liu, Yifan and Hu, Jiancong and Yuan, Yixuan},
  journal={IEEE Journal of Biomedical and Health Informatics},
  volume={26},
  number={10},
  pages={5189--5200},
  year={2022},
  publisher={IEEE}
}

@inproceedings{bai2023llcaps,
  title={Llcaps: Learning to illuminate low-light capsule endoscopy with curved wavelet attention and reverse diffusion},
  author={Bai, Long and Chen, Tong and Wu, Yanan and Wang, An and Islam, Mobarakol and Ren, Hongliang},
  booktitle={International Conference on Medical Image Computing and Computer-Assisted Intervention},
  pages={34--44},
  year={2023},
  organization={Springer}
}

@article{yang2023learning,
  title={Learning to adapt to light},
  author={Yang, Kai-Fu and Cheng, Cheng and Zhao, Shi-Xuan and Yan, Hong-Mei and Zhang, Xian-Shi and Li, Yong-Jie},
  journal={International Journal of Computer Vision},
  volume={131},
  number={4},
  pages={1022--1041},
  year={2023},
  publisher={Springer}
}

@article{zhou2023pyramid,
  title={Pyramid Diffusion Models For Low-light Image Enhancement},
  author={Zhou, Dewei and Yang, Zongxin and Yang, Yi},
  journal={arXiv preprint arXiv:2305.10028},
  year={2023}
}

@article{potlapalli2023promptir,
  title={PromptIR: Prompting for All-in-One Blind Image Restoration},
  author={Potlapalli, Vaishnav and Zamir, Syed Waqas and Khan, Salman and Khan, Fahad Shahbaz},
  journal={arXiv preprint arXiv:2306.13090},
  year={2023}
}

@article{li2023prompt,
  title={Prompt-In-Prompt Learning for Universal Image Restoration},
  author={Li, Zilong and Lei, Yiming and Ma, Chenglong and Zhang, Junping and Shan, Hongming},
  journal={arXiv preprint arXiv:2312.05038},
  year={2023}
}

@article{jiang2023low,
  title={Low-light image enhancement with wavelet-based diffusion models},
  author={Jiang, Hai and Luo, Ao and Fan, Haoqiang and Han, Songchen and Liu, Shuaicheng},
  journal={ACM Transactions on Graphics},
  volume={42},
  number={6},
  pages={1--14},
  year={2023},
  publisher={ACM New York, NY, USA}
}

@inproceedings{garcia2023multi,
  title={Multi-scale structural-aware exposure correction for endoscopic imaging},
  author={Garc{\'\i}a-Vega, Axel and Espinosa, Ricardo and Ram{\'\i}rez-Guzm{\'a}n, Luis and Bazin, Thomas and Falc{\'o}n-Morales, Luis and Ochoa-Ruiz, Gilberto and Lamarque, Dominique and Daul, Christian},
  booktitle={2023 IEEE 20th International Symposium on Biomedical Imaging (ISBI)},
  pages={1--5},
  year={2023},
  organization={IEEE}
}

@InProceedings{li2022air,
    author    = {Li, Boyun and Liu, Xiao and Hu, Peng and Wu, Zhongqin and Lv, Jiancheng and Peng, Xi},
    title     = {All-in-One Image Restoration for Unknown Corruption},
    booktitle = {Proceedings of the IEEE/CVF Conference on Computer Vision and Pattern Recognition (CVPR)},
    month     = {June},
    year      = {2022},
    pages     = {17452-17462}
}

@inproceedings{chen2024lightdiff,
  title={LighTDiff: Surgical endoscopic image low-light enhancement with T-diffusion},
  author={Chen, Tong and Lyu, Qingcheng and Bai, Long and Guo, Erjian and Gao, Huxin and Yang, Xiaoxiao and Ren, Hongliang and Zhou, Luping},
  booktitle={International Conference on Medical Image Computing and Computer-Assisted Intervention},
  pages={369--379},
  year={2024},
  organization={Springer}
}

@inproceedings{bai2024endouic,
  title={Endouic: Promptable diffusion transformer for unified illumination correction in capsule endoscopy},
  author={Bai, Long and Chen, Tong and Tan, Qiaozhi and Nah, Wan Jun and Li, Yanheng and He, Zhicheng and Yuan, Sishen and Chen, Zhen and Wu, Jinlin and Islam, Mobarakol and others},
  booktitle={International Conference on Medical Image Computing and Computer-Assisted Intervention},
  pages={296--306},
  year={2024},
  organization={Springer}
}

@article{wang2024desmoking,
  title={Desmoking of the Endoscopic Surgery Images Based on A Local-Global U-Shaped Transformer Model},
  author={Wang, Wanqing and Liu, Fucheng and Hao, Jianxiong and Yu, Xiangyang and Zhang, Bo and Shi, Chaoyang},
  journal={IEEE Transactions on Medical Robotics and Bionics},
  year={2024},
  publisher={IEEE}
}

@article{su2023multi,
  title={Multi-stages de-smoking model based on CycleGAN for surgical de-smoking},
  author={Su, Xinpei and Wu, Qiuxia},
  journal={International Journal of Machine Learning and Cybernetics},
  volume={14},
  number={11},
  pages={3965--3978},
  year={2023},
  publisher={Springer}
}

@article{pan2022desmoke,
  title={DeSmoke-LAP: improved unpaired image-to-image translation for desmoking in laparoscopic surgery},
  author={Pan, Yirou and Bano, Sophia and Vasconcelos, Francisco and Park, Hyun and Jeong, Taikyeong Ted and Stoyanov, Danail},
  journal={International Journal of Computer Assisted Radiology and Surgery},
  volume={17},
  number={5},
  pages={885--893},
  year={2022},
  publisher={Springer}
}

@inproceedings{ren2024moe,
  title={Moe-diffir: Task-customized diffusion priors for universal compressed image restoration},
  author={Ren, Yulin and Li, Xin and Li, Bingchen and Wang, Xingrui and Guo, Mengxi and Zhao, Shijie and Zhang, Li and Chen, Zhibo},
  booktitle={European Conference on Computer Vision},
  pages={116--134},
  year={2024},
  organization={Springer}
}

@article{ramirez2002detection,
  title={Detection and removal of fat particles from postoperative salvaged blood in orthopedic surgery},
  author={Ram{\'\i}rez, Gemma and Romero, Adolfo and Garc{\'\i}a-Vallejo, Juan Jes{\'u}s and Munoz, Manuel},
  journal={Transfusion},
  volume={42},
  number={1},
  pages={66--75},
  year={2002},
  publisher={Wiley Online Library}
}

@article{ma2021structure,
  title={Structure and illumination constrained GAN for medical image enhancement},
  author={Ma, Yuhui and Liu, Jiang and Liu, Yonghuai and Fu, Huazhu and Hu, Yan and Cheng, Jun and Qi, Hong and Wu, Yufei and Zhang, Jiong and Zhao, Yitian},
  journal={IEEE Transactions on Medical Imaging},
  volume={40},
  number={12},
  pages={3955--3967},
  year={2021},
  publisher={IEEE}
}

@inproceedings{jiang2024autodir,
  title={Autodir: Automatic all-in-one image restoration with latent diffusion},
  author={Jiang, Yitong and Zhang, Zhaoyang and Xue, Tianfan and Gu, Jinwei},
  booktitle={European Conference on Computer Vision},
  pages={340--359},
  year={2024},
  organization={Springer}
}

@article{rao1951asymptotic,
  title={An asymptotic expansion of the distribution of Wilk's criterion},
  author={Rao, C Radhakrishna},
  journal={Bulletin of the international statistical institute},
  volume={33},
  number={2},
  pages={177--180},
  year={1951},
  publisher={International Statistical Institute}
}

@inproceedings{chen2018deeplabv3plus,
  title={Encoder-decoder with atrous separable convolution for semantic image segmentation},
  author={Chen, Liang-Chieh and Zhu, Yukun and Papandreou, George and Schroff, Florian and Adam, Hartwig},
  booktitle={Proceedings of the European conference on computer vision (ECCV)},
  pages={801--818},
  year={2018}
}

@inproceedings{xia2023diffir,
  title={Diffir: Efficient diffusion model for image restoration},
  author={Xia, Bin and Zhang, Yulun and Wang, Shiyin and Wang, Yitong and Wu, Xinglong and Tian, Yapeng and Yang, Wenming and Van Gool, Luc},
  booktitle={Proceedings of the IEEE/CVF International Conference on Computer Vision},
  pages={13095--13105},
  year={2023}
}

@inproceedings{park2023all,
  title={All-in-one image restoration for unknown degradations using adaptive discriminative filters for specific degradations},
  author={Park, Dongwon and Lee, Byung Hyun and Chun, Se Young},
  booktitle={2023 IEEE/CVF Conference on Computer Vision and Pattern Recognition (CVPR)},
  pages={5815--5824},
  year={2023},
  organization={IEEE}
}

@inproceedings{ai2024multimodal,
  title={Multimodal prompt perceiver: Empower adaptiveness generalizability and fidelity for all-in-one image restoration},
  author={Ai, Yuang and Huang, Huaibo and Zhou, Xiaoqiang and Wang, Jiexiang and He, Ran},
  booktitle={Proceedings of the IEEE/CVF Conference on Computer Vision and Pattern Recognition},
  pages={25432--25444},
  year={2024}
}

@inproceedings{qin2024restore,
  title={Restore anything with masks: Leveraging mask image modeling for blind all-in-one image restoration},
  author={Qin, Chu-Jie and Wu, Rui-Qi and Liu, Zikun and Lin, Xin and Guo, Chun-Le and Park, Hyun Hee and Li, Chongyi},
  booktitle={European Conference on Computer Vision},
  pages={364--380},
  year={2024},
  organization={Springer}
}

@inproceedings{tian2025degradation,
  title={Degradation-Aware Feature Perturbation for All-in-One Image Restoration},
  author={Tian, Xiangpeng and Liao, Xiangyu and Liu, Xiao and Li, Meng and Ren, Chao},
  booktitle={Proceedings of the Computer Vision and Pattern Recognition Conference},
  pages={28165--28175},
  year={2025}
}

@inproceedings{luo2025visual,
  title={Visual-Instructed Degradation Diffusion for All-in-One Image Restoration},
  author={Luo, Wenyang and Qin, Haina and Chen, Zewen and Wang, Libin and Zheng, Dandan and Li, Yuming and Liu, Yufan and Li, Bing and Hu, Weiming},
  booktitle={Proceedings of the Computer Vision and Pattern Recognition Conference},
  pages={12764--12777},
  year={2025}
}

@inproceedings{li2024foundir,
  title={Foundir: Unleashing million-scale training data to advance foundation models for image restoration},
  author={Li, Hao and Chen, Xiang and Dong, Jiangxin and Tang, Jinhui and Pan, Jinshan},
  booktitle={Proceedings of the IEEE/CVF international conference on computer vision},
  year={2025}
}

@article{yang2025restore,
  title={Restore-rwkv: Efficient and effective medical image restoration with rwkv},
  author={Yang, Zhiwen and Li, Jiayin and Zhang, Hui and Zhao, Dan and Wei, Bingzheng and Xu, Yan},
  journal={IEEE Journal of Biomedical and Health Informatics},
  year={2025},
  publisher={IEEE}
}

@inproceedings{yang2024all,
  title={All-in-one medical image restoration via task-adaptive routing},
  author={Yang, Zhiwen and Chen, Haowei and Qian, Ziniu and Yi, Yang and Zhang, Hui and Zhao, Dan and Wei, Bingzheng and Xu, Yan},
  booktitle={International Conference on Medical Image Computing and Computer-Assisted Intervention},
  pages={67--77},
  year={2024},
  organization={Springer}
}

@inproceedings{zamfir2025complexity,
  title={Complexity experts are task-discriminative learners for any image restoration},
  author={Zamfir, Eduard and Wu, Zongwei and Mehta, Nancy and Tan, Yuedong and Paudel, Danda Pani and Zhang, Yulun and Timofte, Radu},
  booktitle={Proceedings of the Computer Vision and Pattern Recognition Conference},
  pages={12753--12763},
  year={2025}
}

@article{li2018joint,
  title={Joint medical image fusion, denoising and enhancement via discriminative low-rank sparse dictionaries learning},
  author={Li, Huafeng and He, Xiaoge and Tao, Dapeng and Tang, Yuanyan and Wang, Ruxin},
  journal={Pattern Recognition},
  volume={79},
  pages={130--146},
  year={2018},
  publisher={Elsevier}
}

@inproceedings{liu2023mda,
  title={MDA-SR: multi-level domain adaptation super-resolution for wireless capsule endoscopy images},
  author={Liu, Tianbao and Chen, Zefeiyun and Li, Qingyuan and Wang, Yusi and Zhou, Ke and Xie, Weijie and Fang, Yuxin and Zheng, Kaiyi and Zhao, Zhanpeng and Liu, Side and others},
  booktitle={International Conference on Medical Image Computing and Computer-Assisted Intervention},
  pages={518--527},
  year={2023},
  organization={Springer}
}

@inproceedings{wu2024self,
  title={Self-supervised video desmoking for laparoscopic surgery},
  author={Wu, Renlong and Zhang, Zhilu and Zhang, Shuohao and Gou, Longfei and Chen, Haobin and Zhang, Lei and Chen, Hao and Zuo, Wangmeng},
  booktitle={European Conference on Computer Vision},
  pages={307--324},
  year={2024},
  organization={Springer}
}

@article{chen2025all,
  title={All-in-One Medical Image Restoration with Latent Diffusion-Enhanced Vector-Quantized Codebook Prior},
  author={Chen, Haowei and Yang, Zhiwen and Hou, Haotian and Zhang, Hui and Wei, Bingzheng and Zhou, Gang and Xu, Yan},
  journal={arXiv preprint arXiv:2507.19874},
  year={2025}
}

\end{document}